\documentclass[aps, pra, reprint, superscriptaddress]{revtex4-2}

\usepackage{bibentry,natbib}
\usepackage{latexsym,bm}
\usepackage{graphicx}
\usepackage{rotating}
\usepackage{amsmath}
\usepackage{color}
 \usepackage[dvipsnames]{xcolor}
\usepackage[colorlinks,citecolor=blue]{hyperref}
\usepackage{multirow}
\usepackage{tabularx}
\usepackage{lineno}
\newcolumntype{C}[1]{>{\centering\arraybackslash}p{#1}}

\DeclareGraphicsExtensions{.jpg}

\begin{document}

\title{Probing Multiple Electric Dipole Forbidden Optical Transitions in Highly Charged Nickel Ions}



\affiliation{State Key Laboratory of Magnetic Resonance and Atomic and Molecular Physics, Innovation Academy for Precision Measurement Science and Technology, Chinese Academy of Sciences, Wuhan 430071, China}
\affiliation{Key Laboratory of Atomic Frequency Standards, Innovation Academy for Precision Measurement Science and Technology, Chinese Academy of Sciences, Wuhan 430071, China}
\affiliation{Shanghai EBIT Laboratory, Key Laboratory of Nuclear Physics and Ion-Beam Application (MOE), Institute of Modern Physics, Fudan University, Shanghai 200433, China}
\affiliation{Institute of Applied Physics and Computational Mathematics, Beijing 100088, China}
\affiliation{Department of Physics, University of New Brunswick, Fredericton, New Brunswick, Canada E3B 5A3}
\affiliation{Department of Physics, University of Nevada, Reno, Nevada 89557, USA}
\affiliation{University of Chinese Academy of Sciences, Beijing 100049, China}
\affiliation{Max-Planck-Institut f\"ur Kernphysik, Heidelberg 69117, Germany}

\author{Shi-Yong Liang}
    \thanks{These authors contributed equally to this work.}
    \affiliation{State Key Laboratory of Magnetic Resonance and Atomic and Molecular Physics, Innovation Academy for Precision Measurement Science and Technology, Chinese Academy of Sciences, Wuhan 430071, China}
    \affiliation{Key Laboratory of Atomic Frequency Standards, Innovation Academy for Precision Measurement Science and Technology, Chinese Academy of Sciences, Wuhan 430071, China}
    \affiliation{University of Chinese Academy of Sciences, Beijing 100049, China}
\author{Ting-Xian Zhang}
    \thanks{These authors contributed equally to this work.}
    \affiliation{State Key Laboratory of Magnetic Resonance and Atomic and Molecular Physics, Innovation Academy for Precision Measurement Science and Technology, Chinese Academy of Sciences, Wuhan 430071, China}
    \affiliation{University of Chinese Academy of Sciences, Beijing 100049, China}
\author{Hua Guan}
    \email[Email: ]{guanhua@wipm.ac.cn, xiao\_jun@fudan.edu.cn, cbli@wipm.ac.cn, klgao@wipm.ac.cn}
    \affiliation{State Key Laboratory of Magnetic Resonance and Atomic and Molecular Physics, Innovation Academy for Precision Measurement Science and Technology, Chinese Academy of Sciences, Wuhan 430071, China}
    \affiliation{Key Laboratory of Atomic Frequency Standards, Innovation Academy for Precision Measurement Science and Technology, Chinese Academy of Sciences, Wuhan 430071, China}
\author{Qi-Feng Lu}
    \affiliation{Shanghai EBIT Laboratory, Key Laboratory of Nuclear Physics and Ion-Beam Application (MOE), Institute of Modern Physics, Fudan University, Shanghai 200433, China}
\author{Jun Xiao}
    \email[Email: ]{guanhua@wipm.ac.cn, xiao\_jun@fudan.edu.cn, cbli@wipm.ac.cn, klgao@wipm.ac.cn}
    \affiliation{Shanghai EBIT Laboratory, Key Laboratory of Nuclear Physics and Ion-Beam Application (MOE), Institute of Modern Physics, Fudan University, Shanghai 200433, China}
\author{Shao-Long Chen}
    \affiliation{State Key Laboratory of Magnetic Resonance and Atomic and Molecular Physics, Innovation Academy for Precision Measurement Science and Technology, Chinese Academy of Sciences, Wuhan 430071, China}
    \affiliation{Key Laboratory of Atomic Frequency Standards, Innovation Academy for Precision Measurement Science and Technology, Chinese Academy of Sciences, Wuhan 430071, China}
    \affiliation{Max-Planck-Institut f\"ur Kernphysik, Heidelberg 69117, Germany}
\author{Yao Huang}
    \affiliation{State Key Laboratory of Magnetic Resonance and Atomic and Molecular Physics, Innovation Academy for Precision Measurement Science and Technology, Chinese Academy of Sciences, Wuhan 430071, China}
    \affiliation{Key Laboratory of Atomic Frequency Standards, Innovation Academy for Precision Measurement Science and Technology, Chinese Academy of Sciences, Wuhan 430071, China}

\author{Yong-Hui Zhang}
    \affiliation{State Key Laboratory of Magnetic Resonance and Atomic and Molecular Physics, Innovation Academy for Precision Measurement Science and Technology, Chinese Academy of Sciences, Wuhan 430071, China}
\author{Cheng-Bin Li}
    \email[Email: ]{guanhua@wipm.ac.cn, xiao\_jun@fudan.edu.cn, cbli@wipm.ac.cn, klgao@wipm.ac.cn}
    \affiliation{State Key Laboratory of Magnetic Resonance and Atomic and Molecular Physics, Innovation Academy for Precision Measurement Science and Technology, Chinese Academy of Sciences, Wuhan 430071, China}
\author{Ya-Ming Zou}
    \affiliation{Shanghai EBIT Laboratory, Key Laboratory of Nuclear Physics and Ion-Beam Application (MOE), Institute of Modern Physics, Fudan University, Shanghai 200433, China}
\author{Ji-Guang Li}
    \affiliation{Institute of Applied Physics and Computational Mathematics, Beijing 100088, China}
\author{Zong-Chao Yan}
    \affiliation{Department of Physics, University of New Brunswick, Fredericton, New Brunswick, Canada E3B 5A3}
    \affiliation{State Key Laboratory of Magnetic Resonance and Atomic and Molecular Physics, Innovation Academy for Precision Measurement Science and Technology, Chinese Academy of Sciences, Wuhan 430071, China}
\author{Andrei Derevianko}
    \affiliation{Department of Physics, University of Nevada, Reno, Nevada 89557, USA}
\author{Ming-Sheng Zhan}
    \affiliation{State Key Laboratory of Magnetic Resonance and Atomic and Molecular Physics, Innovation Academy for Precision Measurement Science and Technology, Chinese Academy of Sciences, Wuhan 430071, China}
\author{Ting-Yun Shi}
    \affiliation{State Key Laboratory of Magnetic Resonance and Atomic and Molecular Physics, Innovation Academy for Precision Measurement Science and Technology, Chinese Academy of Sciences, Wuhan 430071, China}
\author{Ke-Lin Gao}
    \email[Email: ]{guanhua@wipm.ac.cn, xiao\_jun@fudan.edu.cn, cbli@wipm.ac.cn, klgao@wipm.ac.cn}
    \affiliation{State Key Laboratory of Magnetic Resonance and Atomic and Molecular Physics, Innovation Academy for Precision Measurement Science and Technology, Chinese Academy of Sciences, Wuhan 430071, China}
    \affiliation{Key Laboratory of Atomic Frequency Standards, Innovation Academy for Precision Measurement Science and Technology, Chinese Academy of Sciences, Wuhan 430071, China}


\date{\today}

\begin{abstract}

Highly charged ions (HCIs) are promising candidates for the next generation of atomic clocks, owing to their tightly bound electron cloud, which significantly suppresses the common environmental disturbances to the quantum oscillator.
Here we propose and pursue an experimental strategy that, while focusing on various HCIs of a single atomic element, keeps the number of candidate clock transitions as large as possible. Following this strategy, we identify four adjacent charge states of nickel HCIs that offer as many as six optical  transitions. Experimentally, we demonstrated the essential capability of producing these ions in the low-energy compact Shanghai-Wuhan Electron Beam Ion Trap. We measured the wavelengths of four magnetic-dipole ($M$1) and one electric-quadrupole ($E$2) clock transitions with an accuracy of several ppm with a novel calibration method; two of these lines were observed and characterized for the first time in  controlled laboratory settings. Compared to the earlier determinations, our measurements improved wavelength accuracy by an order of magnitude. Such measurements are crucial for constraining the range of laser wavelengths for finding the ``needle in a haystack" narrow  lines. In addition, we calculated  frequencies and quality factors, evaluated sensitivity of these six  transitions to the hypothetical variation of the electromagnetic fine structure constant $\alpha$ needed for fundamental physics applications. We argue that all the six  transitions in nickel HCIs offer intrinsic immunity to all common perturbations of quantum oscillators, and one of them has the projected fractional frequency uncertainty down to the remarkable level of 10$^{-19}$.

\end{abstract}


\maketitle


\section{INTRODUCTION}

Quantum metrology of atomic time-keeping has seen dramatic improvements over the past decade with novel applications spanning from chronometric  geodesy~\cite{Chou1630, Bondarescu2012} to fundamental physics, such as dark matter searches~\cite{DerPos14,ArvHuaTil15} and multi-messenger astronomy~\cite{dailey2020ELF.Concept}. 
Currently, optical atomic clocks using neutral atoms or singly charged ions have demonstrated fractional frequency uncertainties  at the level of $10^{-18}$ or even $10^{-19}$~\cite{Brewer2019, Huntemann2016, McGrew2018, Oelker2019}. These uncertainties refer to the ability to protect the quantum oscillator from environmental perturbations, such as stray magnetic and electric fields. As these existing technologies mature, they are reaching the stage where one needs to understand numerous sources of environmental perturbations in greater detail. In some cases, the perturbations cannot be fully eliminated and one needs to devote significant efforts to measuring and characterizing the perturbations; these lead to non-universal systematic corrections to the clock frequency that are specific to experimental realization of the clock.

Novel classes of atomic clocks must start with quantum oscillators that offer a much more improved inherent immunity to environmental perturbations than the more mature technologies.  
One of such systems is the nuclear clock based on the unique property of the $^{229}$Th nucleus -- the existence of a nuclear transition in a laser-accessible range~\cite{Peik2003,Campbell2012}; unfortunately, despite substantial world-wide efforts~\cite{VonDerWense2016,Seiferle2019}, this transition is yet to be observed directly. The suppression of environmental perturbations for the nuclear oscillator comes from the nuclear size being $\sim 10^{4}$ times smaller than the size of a neutral atom. Alternative novel systems are highly charged ions (HCIs)~\cite{Derevianko2012, Safronova2014}. Similar to the nuclear clock,
here the oscillator size is also substantially reduced, owing to the electronic cloud size shrinking as $1/Z$ with the  increasing ion charge $Z$. 
HCIs were proposed as promising candidates for the next generation of  atomic clocks~\cite{Derevianko2012}. In addition, beyond the improved time-keeping, HCIs open intriguing opportunities for probing new physics beyond the standard model of particle physics~\cite{Berengut2010,Safronova2018}.

Compared to the single, yet to be spectroscopically found nuclear transition, there is a plethora of suitable HCIs (see a review~\cite{Kozlov2018} for a sample of proposals). 
A detailed analysis~\cite{Yudin2014} indicates that with certain HCIs, atomic clocks ``can have projected fractional accuracies beyond the $10^{-20}-10^{-21}$ level for all common systematic effects, such as blackbody radiation, Zeeman, ac-Stark, and quadrupolar shifts''. Moreover, compared to the nuclear clock, where the direct observation of the clock transition remains elusive,  the clock transitions in HCIs can be found with conventional spectroscopy or from atomic-structure computations. 
Indeed, here we report spectrographic  measurements  of wavelengths for five clock transitions with an accuracy of several ppm (see Table~\ref{table:wavelength}), setting up the stage for the more accurate laser spectroscopy.

\begin{table*}
    \caption{\label{table:wavelength}
        Observed and calculated wavelengths of magnetic-dipole ($M$1) and electric-quadrupole ($E$2) transitions, 
        where the lines $a$ through $f$ are candidate clock transitions, in nm. 
        }
    \begin{tabular}{cccccccc}
    \hline
    \hline
    \multirow{2}{*}{Line} & \multirow{2}{*}{Ion}& \multirow{2}{*}{Transition}&\multirow{2}{*}{Type}& NIST & \multicolumn{3}{c}{This work}\\
    \cline{6-8}
     & & & & Vacuum & Air (observed) &Vacuum&Theory \\
    \hline
    $a$& Ni$^{11+}$& $3s^23p^5\;\; ^2\!P_{1/2}-{^2\!P}_{3/2}$ &$M$1& 423.2     & 423.104(2)   & 423.223(2)   & 423.0(6)\\
    $b$& Ni$^{12+}$& $3s^23p^4\;\; ^3\!P_{1}-{^3\!P}_{2}$ & $M$1    & 511.724   & 511.570(2)   & 511.713(2)   & 511.8(6)\\
    $c$ & Ni$^{14+}$& $3s^23p^2\;\; ^3\!P_{1}-{^3\!P}_{0}$ & $M$1   & 670.36    & 670.167(2)   & 670.352(2)   & 671.1(14)\\
    $d$& Ni$^{15+}$& $3s^23p\;\; ^2\!P_{3/2}-{^2\!P}_{1/2}$ & $M$1  & 360.22    & 360.105(2)   & 360.207(2)   & 359.9(9)\\
    $e$& Ni$^{12+}$& $3s^23p^4\;\; ^3\!P_{0}-{^3\!P}_{2}$ & $E$2    & 498.50(249)   & --            & --            & 496.9(24)\\
    $f$& Ni$^{14+}$& $3s^23p^2\;\; ^3\!P_{2}-{^3\!P}_{0}$ & $E$2    & 365.277  & --            & 365.278(1)            & 365.0(3)\\
    $g$ & Ni$^{14+}$& $3s^23p^2\;\; ^3\!P_{2}-{^3\!P}_{1}$ & $M$1   & 802.63   & 802.419(2)    & 802.639(2) & 800.3(25)\\
    \hline
    \hline
    \end{tabular}
\end{table*}

Despite the lack of suitable electric-dipole ($E$1) transitions for direct laser cooling, recent successes in sympathetic cooling and quantum logic spectroscopy of HCIs have paved way for precision spectroscopic measurements with HCIs~\cite{Micke2020, Schmoger2015}. It is worth emphasizing that these newly demonstrated technologies can be applied universally to a wide range of HCIs. The multitude  of suitable  clock HCI candidates is a ``blessing in disguise'', as one needs to commit to building the infrastructure for a specific ion. As with any new endeavor, one would like to mitigate potential problems with picking 
a ``wrong'' ion.  Here we propose  and pursue a straddling strategy that would allow one to explore several clock transitions using not only  the same HCI production system but also ions of the same atomic element.

\begin{figure}
\includegraphics{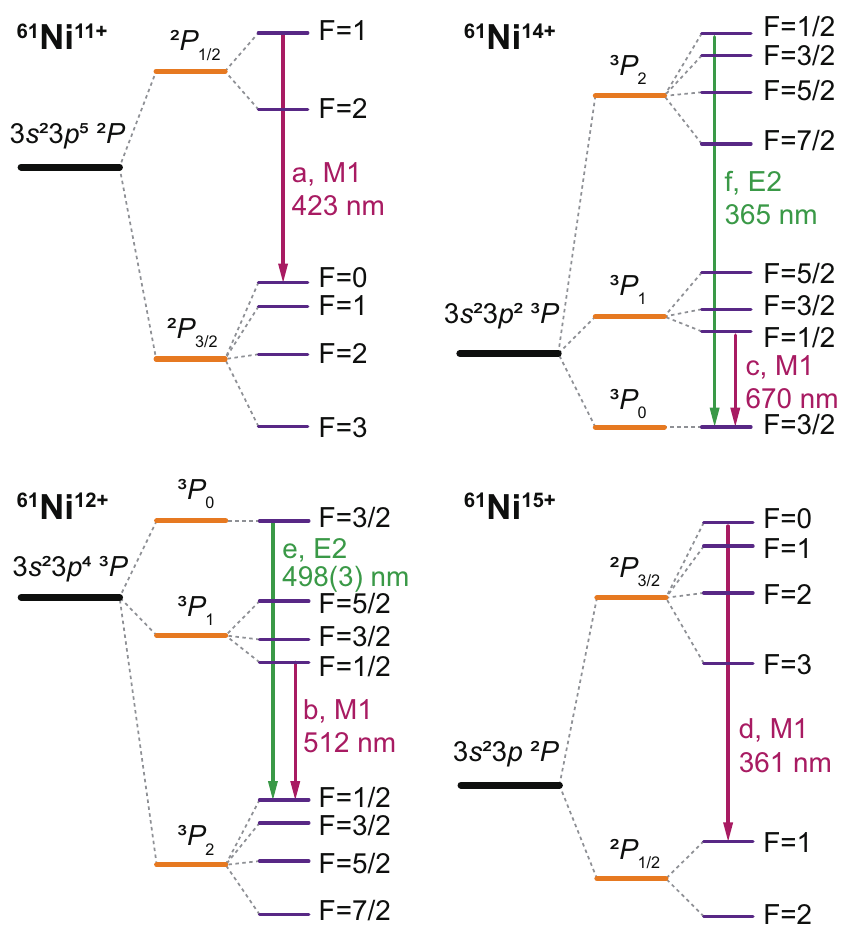}
\caption{\label{Fig:ClockTransions} 
  Partial energy-level diagrams for  highly charged nickel ions. Clock transitions are explicitly drawn. Magnetic-dipole ($M$1) transitions are shown in magenta and electric-quadrupole ($E$2) transitions in green. The labeling of transitions is the same as in Table~\ref{table:wavelength}.}
\end{figure}

A suitable HCI has to possess a number of  properties  enabling precision spectroscopy and  compatibility with operating an atomic clock. Generally, one may distinguish between three classes of visible or near visible optical forbidden transitions in HCIs that can be used for developing optical clocks:
\begin{enumerate}
\item{ Magnetic-dipole ($M$1) transitions between two hyperfine-structure levels of the same electronic state~\cite{Schiller2007,Yudin2014}. }
\item{ Forbidden transitions between level crossing electronic states, which tend to be sensitive to variation of the fine structure
constant~\cite{Berengut2010, Berengut2011, Bekker2019, Dzuba2012}. }
\item{ Forbidden transitions between the ground-state fine structure levels~\cite{Derevianko2012, Yudin2014, Yu2016,Yu2018}. }
\end{enumerate}
Type 1 transitions occur in few-electron heavy HCIs~\cite{Yudin2014} that are challenging to produce and trap.
Type 2 transitions involve a complex energy structure that can impede initialization and read-out of the clock states. Here we focus on type 3 transitions that offer simplicity in both producing the ions and clock operation. More specifically we choose HCIs of nickel of various charge states~\cite{Yudin2014, Yu2016, Yu2018}: Ni$^{11+}$, Ni$^{12+}$, Ni$^{14+}$, and Ni$^{15+}$. The clock transitions are shown in Fig.~\ref{Fig:ClockTransions}.
All the traditional clock  perturbations are strongly suppressed for these ions due to the charge scaling arguments~\cite{BerDzuFla12, Derevianko2012, Yudin2014}. As pointed out in Ref.~\cite{Derevianko2012},
the major issue with HCI clocks is the so-called quadrupolar shift of the clock transition, when the quadrupole ($Q$) moment of the clock state couples to the always existing $E$-field gradients in ion traps. While the $Q$-moment of an electronic cloud does scale as $1/Z^2$, this reduction is not sufficient to suppress the quadrupolar shift below the desired level of accuracy. Thus, it is beneficial to select clock states with either vanishing or strongly suppressed $Q$-moments.

There are four $M$1 and two $E$2 optical transitions in Ni$^{11+}$, Ni$^{12+}$, Ni$^{14+}$, and Ni$^{15+}$ that offer the desired flexibility. These ions have  varying  number of electrons in the $3p$
shell, see Fig.~\ref{Fig:ClockTransions}. The clock transitions are between the fine structure components of the ground electronic state.  
There are four stable isotopes $^{58}$Ni, $^{60}$Ni, $^{62}$Ni, and $^{64}$Ni without nuclear spin; these  can be used to search for new physics with isotope shift measurements~\cite{Counts2020, Solaro2020,BerengutArxiv2020} and for initial spectroscopic measurements. These spin-0 isotopes, however, will be susceptible to the  quadrupolar shifts for clock transitions.
However, these shifts can be suppressed by using
 the  $^{61}$Ni isotope (nuclear spin $I=3/2$), which has a natural abundance of 1.14\%. Then the quadrupolar shifts can be either strongly suppressed or completely removed 
by employing the following clock transitions between hyperfine states (see Fig.~\ref{Fig:ClockTransions}):
\begin{itemize}
\item{${^2\!P}_{3/2} \, F=0$ and $^2\!P_{1/2} \, F=1$ or $F=2$ for Ni$^{11+}$ and Ni$^{15+}$,}
\item{$^3\!P_{1} \, F=1/2$ and $^3\!P_{2} \,  F=1/2$ for Ni$^{12+}$,}
\item{$^3\!P_{1} \, F=1/2$ and $^3\!P_{0} \, F=3/2$ for Ni$^{14+}$,}
\item{$^3\!P_{0} \, F=3/2$ and $^3\!P_{2} \,  F=1/2$ for Ni$^{12+}$ and Ni$^{14+}$.}
\end{itemize}
This selection is based on the following reasoning~\cite{Yudin2014}: $Q$-moments (rank 2 tensors) of  the $F=0$ and $F=1/2$ states  are zero due to selection rules. For the $^2\!P_{1/2} \, F=1,2$ and  $^3\!P_{0} \, F=3/2$ states, the electronic $Q$-moments vanish due to selection rules for the electronic angular momentum $J$. Thereby, the $Q$-moments are determined by the nuclear $Q$-moments or hyperfine mixing~\cite{Campbell2012} and, as such, are strongly suppressed.
As an indication of attainable accuracy, Refs.~\cite{Yu2016,Yu2018} evaluated relevant properties of the clock transitions in $^{61}$Ni$^{15+}$ and  $^{58}$Ni$^{12+}$ and  
estimated common systematic uncertainties to be below $10^{-19}$,
in line with the more general estimates of Ref.~\cite{Yudin2014}. 
The second-order Doppler shift induced by the excess micromotion of the trapped ion is expected to be suppressed to below $10^{-19}$ by compensating the stray electric field to a level below 0.1~V/m~\cite{Huber2014, Huang2019}. In a cryogenic trap, the heating rate of the trapped ions caused by the collisions with the background gas and the anomalous motional heating is reduced, and hence the second-order Doppler shift induced by the secular motion is also expected to be sufficiently small~\cite{Kozlov2018}.
Based on these arguments, we expect the attainable fractional systematic uncertainty 
of all the six clock transitions in Ni HCIs to be  $10^{-19}$.

As the first essential step towards realizing the Ni HCI clocks, we produced the target ions  at our newly built low-energy compact Shanghai-Wuhan Electron Beam Ion Trap (SW-EBIT)~\cite{Liang2019}.
The wavelengths of four $M$1 and one $E$2 clock transitions between the ground-state fine structure levels in these ions are measured to an accuracy of several ppm using a spectrograph.
In particular, the three $M$1 lines $b$, $c$, and $g$ (listed in Table~\ref{table:wavelength}) in Ni$^{12+}$ and Ni$^{14+}$ are observed and characterized for the first time in the laboratory.
We also carried out calculations for these ions using an {\em ab initio} relativistic method of atomic structure,  the 
multi-configuration Dirac-Hartree-Fock (MCDHF) method~\cite{Grant2007,Fischer2016}. We evaluated relevant spectroscopic properties, such as transition wavelengths and natural linewidths. 
We also estimated the sensitivity to the hypothetical variation of the fine structure constant $\alpha$ and found that all considered clock transitions in Ni HCIs are more susceptible to the variation than most of the 
commonly employed singly charge ions or neutral atoms. Thus, Ni HCIs 
can be used for placing stringent constraints on the spatial or temporal  variation of $\alpha$.

\section{EXPERIMENTAL METHOD AND RESULTS}
\subsection{Production of Ni HCIs}

\begin{figure}
\includegraphics{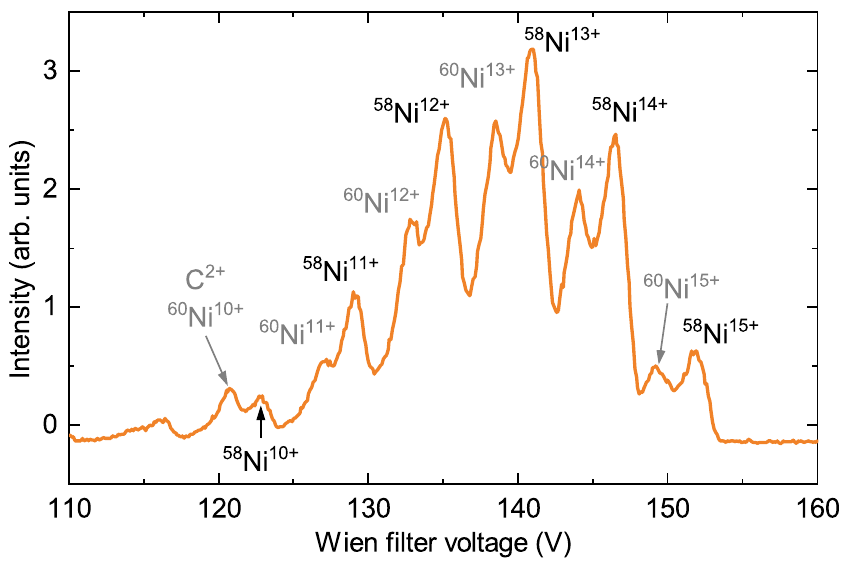}
\caption{\label{fig:chargeState}
    Charge-state distribution of the Ni HCIs, obtained by averaging 3 measurements. 
}
\end{figure}

To produce Ni HCIs, we injected the Ni(C$_5$H$_5$)$_2$ (nickelocene) molecular beam into the trap center. Then the charge-state distribution of Ni HCIs was measured using the electron-beam current of 6~mA and the electron-beam energy of 500~eV, which is higher than the ionization energies 319.5~eV, 351.6~eV, 429.3~eV, and 462.8~eV needed for Ni$^{11+}$, Ni$^{12+}$, Ni$^{14+}$, and Ni$^{15+}$, respectively. The extraction period was 0.3~s and the  magnetic flux density was 0.16~T. 
As shown in Fig.~\ref{fig:chargeState}, the target ions Ni$^{11+}$, Ni$^{12+}$, Ni$^{14+}$, and Ni$^{15+}$ were produced, and the ions of two distinct isotopes,  $^{60}$Ni and $^{58}$Ni, were resolved.
The techniques for measuring charge-state distribution are described in Ref.~\cite{Liang2019}.

\subsection{\label{sec:spectra} Spectral measurements}

\begin{figure}
\includegraphics{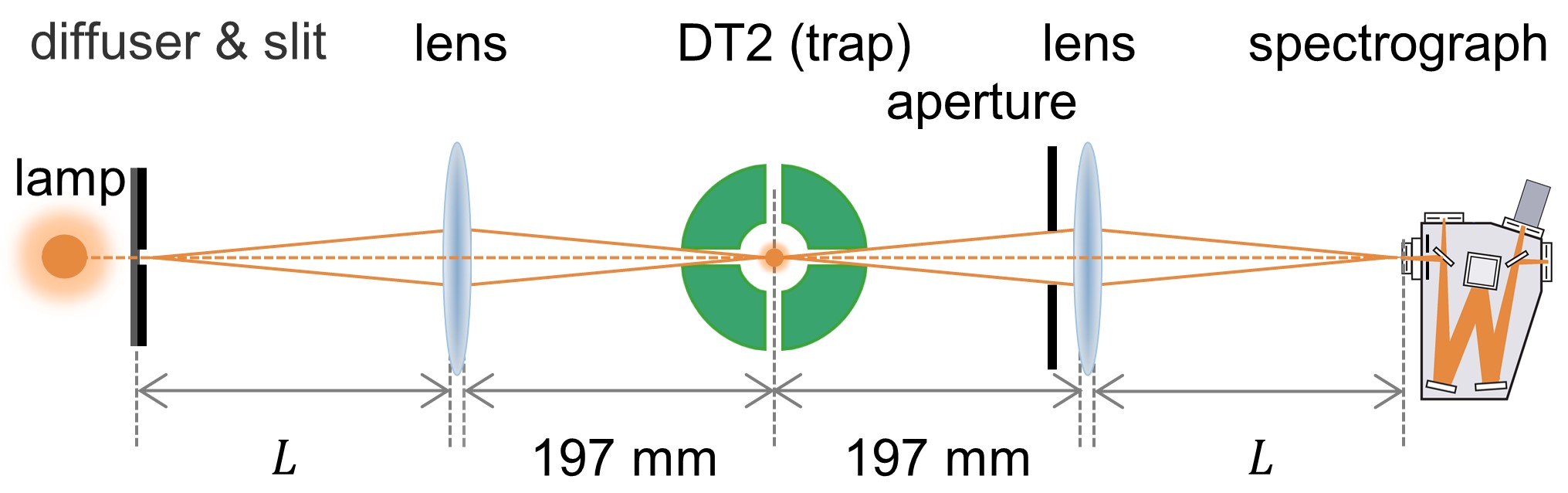}
    \caption{\label{fig:scheme}
        Scheme of observation and calibration of the measured lines. The Ni HCIs are trapped at the center of DT2 in SW-EBIT.
    }
\end{figure}

The spectra of the trapped Ni HCIs were observed by a Czerny-Turner spectrograph (Andor Kymera 328i) equipped with an Electron Multiplying Charge-Coupled Device (EMCCD, Andor Newton 970, pixel: $1600 \times 200$, pixel size: 16 $\mu $m) and a 1200 l/mm grating blazed at 500~nm.
To maximize the number of the Ni HCIs of a specific charge state, different electron-beam energies were used, \textit{i.e.} 370~eV, 400~eV, 500~eV, and 540~eV for Ni$^{11+}$, Ni$^{12+}$, Ni$^{14+}$, and Ni$^{15+}$, respectively.
As illustrated in Fig.~\ref{fig:scheme}, the fluorescence emitted from the Ni HCIs was focused by a single N-BK7 Bi-Convex lens (focal length $f=10$~cm at 633~nm) on the spectrograph entrance slit.
The distance between the trap (DT2, drift tube~2 in SW-EBIT~\cite{Liang2019}) center and the front principal plane of the lens remained fixed at 197~mm, which was about twice the focal length.
Before setting up the spectrograph, a Charge-Coupled Device (CCD) was placed on the image plane to image the two inner edges of DT2 (1~mm slit width) that were illuminated by the hot cathode.
In order to ensure that the lens was aligned with the optical axis, we adjusted the angle and position of the lens until the edge image became mirror-symmetric.
Because of the dispersion of the lens, to ensure that the spectrograph slit was always precisely located on the image plane, the distance $L$ between the slit of the spectrograph and the back principal plane of the lens was calculated and adjusted for every central wavelength of the measured spectra.
The grating was set to zero-order to image the inner edges of DT2 through spectrograph with its maximum slit width and minimum iris aperture behind the slit. 
Similarly, the angle and position of the spectrograph were adjusted until the image of the edges became mirror-symmetric to ensure the spectrograph alignment  with the optical axis.
A one inch aperture was placed before the lens to block the stray light.
For calibration, a conjugated optical system was installed on the opposite side of the spectrograph.
A diffuser attached by a 0.5~mm slit was placed on the object plane.
A low-pressure gas-discharge lamp filled with Kr illuminated the slit, and the slit was imaged to the trap center to overlap with the trapped ion cloud. 
During the spectral exposure time of 10 to 60 minutes, the Kr lamp as the calibration light source flashed at a period of 1 to 3 minutes.
The slit of the spectrograph was set to 30~$\mu$m, and the iris aperture in the spectrograph was set to 40 steps to obtain the F/7.6 aperture.
The focal length of the spectrograph was tuned to minimize the linewidth in each spectral range. 

\begin{figure*}
\includegraphics{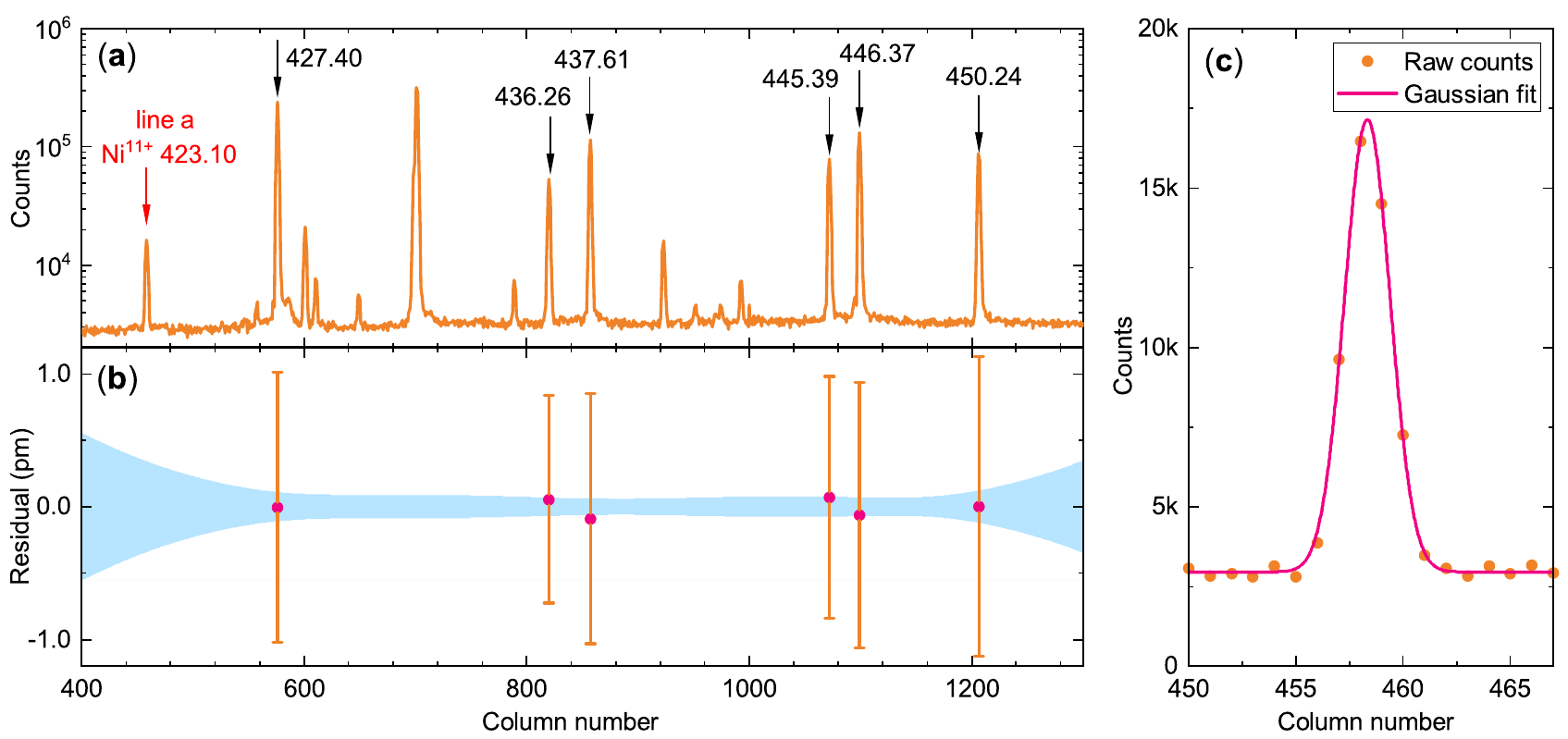}
    \caption{\label{fig:spectrum}
        (a) A spectrum of line $a$ from Ni$^{11+}$ and its calibration lines from Kr atom whose approximate wavelengths are labeled in the figure in nm. (b) Residuals of cubic polynomial fits of the calibration lines. The gray band is a 1-$\sigma$ confidence band. The uncertainties in the calibration lines are dominated by the line centroid uncertainties of the Gaussian fits. (c) Spectrum of line $a$ and its Gaussian fit. }
\end{figure*}

All the spectra were binned to a non-dispersive direction after removing the cosmic ray noise, as shown in Fig.~\ref{fig:spectrum}~(a).
The dispersion function was obtained by fitting the NIST-tabulated Ritz in-the-air wavelengths of the calibration lines to a cubic polynomial, against their column numbers of the line centroids.
The residuals of the calibration lines and the 1-$\sigma$ fitting confidence band are shown in Fig.~\ref{fig:spectrum}~(b).
To determine the line centroids, the measured lines and calibration lines were fitted to a Gaussian or a multi-Gaussian profile, as shown in Fig.~\ref{fig:spectrum}~(c).

\subsection{Observed wavelengths}

Previously, these five $M$1 lines
 in Table~\ref{table:wavelength}
were observed in the solar corona emission~\cite{Jefferies1971} with a wavelength uncertainty of  tens of picometer.
The lines $a$ and $d$ have been also measured in Tokamak~\cite{Behringer1986, Hinnov1990}, but experimental observation of the other three lines $b$, $c$, and $g$ has not been reported in the literature.
In this work, we observed and identified all five $M$1 lines emitted from the nickel plasma in the SW-EBIT in a controlled laboratory setting.
The measured wavelengths agree with the Ritz wavelengths in NIST database~\cite{NIST_ASD}, as shown in Table~\ref{table:wavelength}, where the wavelengths between air and vacuum were converted by an empirical equation~\cite{Peck1972}.
However, for line $d$, our result of 360.105(2)~nm  substantially deviates from the value of 360.123(2)~nm observed from Tokamak plasma by \citet{Hinnov1990}.
To test our result, two lines from Ar$^+$ were measured without any change of the optical system comparing to the measurement of line $d$, and the measured wavelengths in air were  357.660(2)~nm and 358.843(2)~nm, which were in good agreement with the Ritz wavelengths in NIST database, \textit{i.e.}, 357.661538~nm and 358.844021~nm.
For the $E$2 lines $e$ and $f$, the transition rates are too small to be observable by our technique. However, we deduced the wavelength of line $f$ in Table~\ref{table:wavelength}  from those of lines $c$ and  $g$ via the Rydberg-Ritz combination principle.

\subsection{Measurement uncertainties}

\begin{figure}
\includegraphics{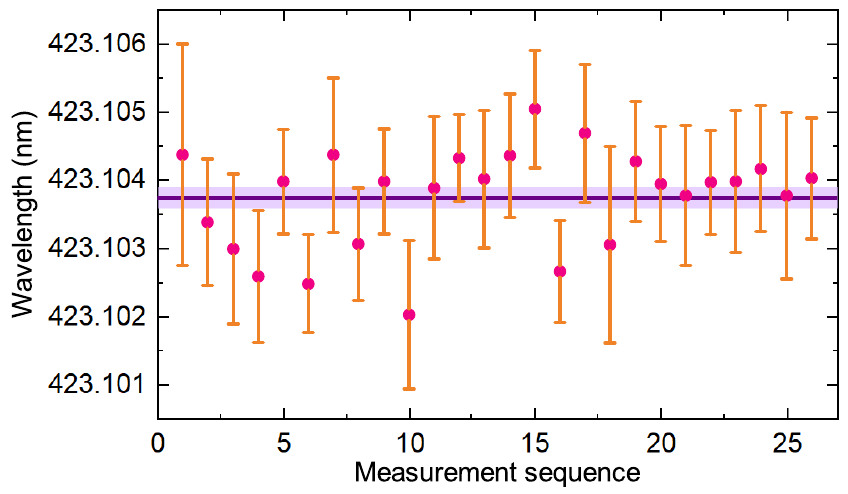}
    \caption{\label{fig:wavelength}
    The calibrated wavelengths of line $ a $ in air derived from a series of 26 measurements. The wavelength uncertainty of each single spectrum was calculated as the quadrature of the line centroid uncertainty and the 1-$\sigma$ confidence interval of the fitted dispersion function.
    The weighted average wavelength is represented by the solid purple line and its uncertainty is represented by lilac band.
    }
\end{figure}

\textbf{Line centroid uncertainty.}
The line centroids of the measured lines and their calibration lines were determined by the centers of the fitted Gaussian profiles.
Since the statistical uncertainty of the line centroid
was mainly caused by the low signal-to-noise ratio, we evaluated the statistical uncertainty by performing at least 20 measurements on the line of interest,
as shown in Fig~\ref{fig:wavelength}.
For all five measured lines, this uncertainty was smaller than 0.4~pm. 
The systematic uncertainty of the line centroid is mainly caused by the non-ideal Gaussianity of the line because of the optical aberration and the Zeeman components.
In this work, since the measured lines and their calibration lines shared a similar profile, the optical aberration effect was largely offset.
In the trap center, the magnetic flux density was $\sim$0.16~T, resulting in a $\sim$2~pm splitting between the Zeeman components of the clock transitions, which was relatively small (unresolved) compared to the $\sim$90~pm linewidth.
Furthermore, the Zeeman effect would not alter the line centroid because
the Zeeman components were symmetrically distributed; in addition, the Zeeman effect was negligible for the Kr lamp due to the low magnetic field of 0.4~mT.

\textbf{Dispersion function uncertainty.}
The statistical uncertainty for the dispersion function was caused by the centroid statistical uncertainties of calibration lines, which were reduced by the statistics of the line centroids.
The systematic dispersion function uncertainty of a measured line was estimated by averaging the absolute values of the fitted residuals of its calibration lines of all the measured spectra.

\textbf{Calibration systematic uncertainty.}
Since the image of the calibration light source might not be overlapped exactly with the trapped ion cloud, the spatial deviation and misalignment  could cause wavelength offset between the measured lines and their calibration lines. In this work, a spatial deviation of less than 2~mm would result in a wavelength uncertainty of less than 1~pm.
The misalignment could cause a wavelength uncertainty of less than 1~pm, which was
estimated from five measurements of the  Ar$^{9+}$ 553~nm line by resetting the optical system every time.
Thereby, the overall systematic uncertainty caused by our calibration scheme was expected to be less than 2~pm.

\textbf{Other uncertainties.}
In this work, the calibration light source and the fluorescence of the trapped ions were exposed to the spectrograph almost simultaneously, indicating that the temperature drift and the mechanical drift were canceled out.
The shifts due to the Stark effect and collisions
can also be neglected at this level of accuracy. The wavelengths of the selected calibration lines are reliable because their uncertainties in the NIST database are all less than 0.3~pm.

Table~\ref{table:uncertainty} is the uncertainty budget for the lines $a$-$d$ and $g$ in air. The total uncertainty was calculated as the quadrature of all the uncertainties, which was dominated by the calibration systematic uncertainty.
In order to test the reliability of the uncertainty estimation, the wavelengths in-the-air of the three lines from Ar HCI were measured,
\textit{i.e.}, Ar$^{9+}$ 553.327(2)~nm, Ar$^{10+}$ 691.689(2)~nm, and Ar$^{13+}$ 441.255(2)~nm, which were consistent with the previous measured values of 553.3265(2)~nm, 691.6878(12)~nm, and 441.255919(6)~nm, respectively~\cite{Draganic2003, Egl2019}.

The total wavelength uncertainty of this observation and calibration scheme was approximately 2~pm, which was comparable to the uncertainty of the scheme that the measured lines were calibrated by the lines from the buffer gas observed by a similar resolution spectrograph~\cite{Kimura2019-2, Kimura2019}, but larger than the uncertainty of the scheme that the calibration source was overlapped with the real image of the ion cloud observed by a higher resolution spectrograph~\cite{Draganic2003}. Compared to these two schemes, our scheme is more convenient and flexible. The uncertainty may be reduced by using a higher resolution spectrograph that is less sensitive to the calibration optical system.

\begin{table}
\caption{
    \label{table:uncertainty}
    Uncertainty budget of the measured lines. 
    }
\begin{tabular}{lC{9mm}C{9mm}C{9mm}C{9mm}C{9mm}}
\hline \hline
Source & \multicolumn{5}{c}{Uncertainty in wavelength (pm)}  \\
\hline
Line                               & $a$       &  $b$    & $c$    & $d$  & $g$   \\
Line centroid         & 0.2        & 0.2       & 0.1     & 0.2    & 0.3  \\
Dispersion function   & 0.1        & 0.3       & 0.3     & 0.2    & 0.4       \\
Calibration systematic & 2       & 2      & 2    & 2   & 2     \\
Total                              & 2          & 2         & 2       & 2      & 2        \\ \hline \hline
\end{tabular}
\end{table}

\section{THEORETICAL METHOD AND RESULTS}

\subsection{ MCDHF calculations}

In the MCDHF method, an atomic wave function $\Psi$ is constructed as a linear combination  of configuration state functions (CSFs) $\Phi$ of the same parity $P$, the total angular momentum $J$, and its projection $M_J$, {\it i.e.},
\begin{equation}
    \Psi(\Gamma PJM_J) = \sum_{i=1}^{N_\mathrm{CSF}}{c_i \Phi (\gamma_iPJM_J)}.
\end{equation}
Here $c_i$ is the mixing coefficient and $\gamma _i$ stands for the remaining quantum numbers of the CSFs.
Each CSF itself is a linear combination of products of one-electron Dirac orbitals. Both mixing coefficients and orbitals are optimized in the self-consistent field calculation.
After a set of orbitals is obtained, the relativistic configuration interaction (RCI) calculations are used to capture more electron correlations.
In addition to the Coulomb interactions, our RCI calculations also include the
Breit interaction in the low-frequency approximation 
and the quantum electrodynamic (QED) corrections.

In order to obtain high-quality atomic wave functions, we designed an elaborate computational model as follows. In the first step, the self-consistent field (SCF) calculations were performed successively to generate virtual orbitals. The virtual orbitals were employed to form CSFs which account for certain electron correlations. More specifically, CSFs were produced through single (S)- and double (D)-electron excitations from the occupied Dirac-Hartree-Fock orbitals to virtual orbitals, but the double excitation from the atomic core $1s^22s^22p^6$ orbitals were excluded at this stage. The virtual orbitals were augmented layer by layer up to $n_{\rm{max}}=12$ and $l_{\rm{max}}=6$, where $n_{\rm{max}}$ and $l_{\rm{max}}$ denote, respectively, the maximum principal quantum number and the maximum angular quantum number of the virtual orbitals. 
In the second step, the single-reference configuration RCI calculations were performed with the configuration space constructed from SD excitation from all occupied orbitals to the set of virtual orbitals. In other words, the correlation between electrons in the atomic core, which were neglected in the first step, were captured. In the last step, we considered part of contributions from the triple- and quadruple-excitation CSFs. In order to limit the number of CSFs, the MR-SD approach was adopted to produce corresponding CSFs~\cite{Li2012,Li2016}. 
The multi-reference (MR) configuration sets were created as $\{3s 3p^5 3d$, $3s^2 3p^3 3d^2\}$  for Ni$^{11+}$, $\{3s3 p^4 3d$, $3s^2 3p^2 3d^2$, $3p^6\}$ for Ni$^{12+}$, $\{3s 3p^2 3d$, $3s^2 3d^2$, $3p^4\}$ for Ni$^{14+}$, and $\{3p^3$, $3s 3p 3d$, $3p 3d^2\}$ for Ni$^{15+}$.
Additionally, the Breit interaction and the QED corrections were included in the RCI computation. 

Once the atomic wave functions are obtained, we are in a position to calculate the physical quantities under investigation, that is, the reduced matrix elements for corresponding rank $k$ irreducible tensor operators between 
 two atomic states, {\it i.e.},
$
    \langle \Psi (\Gamma PJ) \|O^{(k)}\| \Psi (\Gamma^{'} P^{'}J^{'})\rangle \,.
$
The magnetic dipole and electric quadrupole transition operators are rank 1 and rank 2 operators, respectively. 
In practice, we performed the calculations using the GRASP2018 package~\cite{FroeseFischer2019}.

\subsection{Calculated wavelengths}

As shown in Table~\ref{table:wavelength}, the calculated wavelengths of the $M$1 transitions  of line $a$ through line $d$ and line $g$ agree with our measured values.  The wavelengths of the two $E$2 transitions of line $e$ and line $f$ in Ni$^{12+}$ and Ni$^{14+}$ were also calculated. These two  lines have not  been observed yet before. Our calculated wavelengths for these two transitions are in agreement with the NIST recommended values. Meanwhile, the calculated wavelength of line $f$ also agrees with our indirect measurement, see Table~\ref{table:wavelength}.

\subsection{Properties of the clock transitions}
The design of an atomic clock relies on the knowledge of atomic parameters of the quantum oscillator.
Thus, we have computed wavelengths, spontaneous emission rates $A$, lifetimes $\tau$,  linewidths $\Gamma$~($2\pi \Gamma= 1/\tau$), and other parameters for all six candidate clock transitions, and the results are listed in Table~\ref{table:atomicpar}. As one of the key parameters of clock stability, the quality factor ($Q$-factor) is also given in this table. The $Q$-factor is defined as the ratio of the clock frequency $\nu_\mathrm{clk}$ to the linewidth $\Gamma$ of the clock transition, {\it i.e.}, $Q=\nu_\mathrm{clk}/\Gamma$. 
Among the four $M$1 clock transitions, the $^3\!P_1-{^3\!P}_0$ transition in Ni$^{14+}$ is the narrowest with its linewidth less than
$10\,\mathrm{Hz}$, while the linewidths of the other three $M$1 transitions are about $30\,\mathrm{Hz}$. The corresponding $Q$-factors of these four $M$1 transitions are $\sim 10^{13}$. 
There are two decay channels from $^3\!P_0$ in Ni$^{12+}$ and $^3\!P_2$ in Ni$^{14+}$ to the lower states. In order to determine the linewidth of these $E$2-clock transitions , both decay channels should be taken into account. For  $^3\!P_0$ in Ni$^{12+}$, the decay rate is 0.037~s$^{-1}$ for the 
$E$2~($^3\!P_0-{^3\!P}_2$) channel and 0.011~s$^{-1}$ for the 
$M$1~($^3\!P_0-{^3\!P}_1$) channel. For $^3\!P_2$ of Ni$^{14+}$, the 
$E$2~($^3\!P_2-{^3\!P}_0$) and $M$1~($^3\!P_2-{^3\!P}_1$) transition rates are 
0.03~s$^{-1}$ and 22.5~s$^{-1}$, respectively. Therefore, the linewidths for the $E$2-clock transitions are $3.6~\mathrm{Hz}$ for $^3\!P_2-{^3\!P}_0$ in Ni$^{14+}$ and $8~\mathrm{mHz}$ for $^3\!P_0-{^3\!P}_2$ in Ni$^{12+}$, which are respectively
smaller than the $M$1 transition lines {\it c} and {\it b}, as marked in Fig.~\ref{Fig:ClockTransions}. This $E$2 transition in Ni$^{12+}$ is particularly attractive for stable clockwork~\cite{Yu2018}, because of its relatively high $Q$-factor of $7.5 \times 10^{16}$, meaning that the statistical uncertainty limited by the quantum projection noise~\cite{Itano1993, Peik2006, Kozlov2018} of this transition can reach the level of $10^{-19}$ by averaging over a few days.

\begin{table*}
    \caption{\label{table:atomicpar}
  Theoretical spectral properties of clock transitions. 
 Here $A$ is the Einstein coefficient for spontaneous decay, $\tau$ is the lifetime of the upper clock state, $\Gamma$ is the natural linewidth, and $Q$ is the transition quality factor. Also, $q$ and $K$ are, respectively, the sensitivity coefficient and enhancement factor for the variation of the fine structure constant. Numbers in square brackets stand for the powers of 10, {\it i.e.},  $x[y] \equiv x \times 10^y$. 
      }
     \setlength{\tabcolsep}{2mm}{
    \begin{tabular}{ccccccccccc}
     \hline
     \hline
 Transition                                            & Type & $A$ (s$^{-1}$) & $\tau$ (ms)  & $\Gamma$ (Hz) & $Q$     & $q$ (cm$^{-1}$)   & $K$ & \\\hline
                                                        \multicolumn{9}{c}{Ni$^{11+}$  $3s^23p^5$}\\
  $^2\!P_{1/2}-{^2\!P}_{3/2}$&  $M$1  & 238(2)         & 4.2(1)       & 38            & 1.9[13] & 24820 & 2.1 & \\
                                             &      & 236.31(3)      & 4.23(2)      &               &         & 24464 &     & Ref.~\cite{Nandy2013}    \\
                                                        \multicolumn{9}{c}{Ni$^{12+}$ $3s^23p^4$}\\
  $^3\!P_{1}-{^3\!P}_{2}$                    &  $M$1  & 157(1)         & 6.3(1)       & 25            & 2.3[13] & 22473 & 2.3 & \\                                                             &      & 154            & 6.5          &               &         &       &     & Ref.~\cite{Yu2018}    \\
    $^3\!P_{1}-{^3\!P}_{2}$                    &  $E$2  & 0.02         &         &             &  &   &  & \\ 
  $^3\!P_{0}-{^3\!P}_{2}$                    &  $E$2  & 0.037(4)       & 21(3)[3]      &0.008         &  7.5[16] & 14982 & 1.5 & \\    
                                             &      & 0.03           & 19[3]        &               & 1.1[16] &       &     & Ref.~\cite{Yu2018} \\
  $^3\!P_{0}-{^3\!P}_{1}$                    &  $M$1  & 0.011(2)        &              &               &         &       &     & \\
                                                        \multicolumn{9}{c}{Ni$^{14+}$ $3s^23p^2$}\\
  $^3\!P_{1}-{^3\!P}_{0}$                    &  $M$1  & 56.1(5)        & 17.8(2)      & 9             & 5.0[13] & 20340 & 2.7 & \\
  $^3\!P_{2}-{^3\!P}_{0}$                    &  $E$2  & 0.030(1)      & 44(1)        & 3.6           & 2.3[14] & 28197 & 2.1 & \\
  $^3\!P_{2}-{^3\!P}_{1}$                    &  $M$1  & 22.5(4)         &              &               &         &       &     & \\
    $^3\!P_{2}-{^3\!P}_{1}$                  &  $E$2  & 0.001         &              &               &         &       &     & \\
                                                        \multicolumn{9}{c}{Ni$^{15+}$ $3s^23p$}\\
  $^2\!P_{3/2}-{^2\!P}_{1/2}$  &  $M$1  & 193(2)         & 5.2(1)       & 31            & 2.7[13] & 29204 & 2.1 & \\
                                                          &      & 190.99         & 5            & 30.38         & 2.73[13]& 89391 &     & Ref.~\cite{Yu2016} \\ 
                                                         
 \hline
 \hline
 \end{tabular}}
 \end{table*}

From the perspective of searching for new physics, we anticipate that by monitoring the Ni HCI clock transition frequencies, stringent constraints could be placed on the possible time variation of the fine structure constant $\alpha$. Following Refs.~\cite{Dzuba1999,Safronova2018}, one can introduce the ``sensitivity coefficient'' $q$, defined by $\omega(x)=\omega_0 + q x$, where $x \equiv (\alpha / \alpha_0)^2 - 1$ and $\omega_0$ is the clock transition frequency at the nominal value of the fine structure constant $\alpha_0$. The sensitivity coefficient $q$ characterizes the linear response of the clock frequency $\omega(x)$ to the variation of $\alpha$, and can be 
calculated numerically as $q\approx [\omega(+x) - \omega(-x)]/{(2x)}$.
Another commonly used quantity is   
the dimensionless enhancement factor~\cite{Safronova2018} $K= \partial \ln \omega/\partial \ln \alpha \approx  2q / \omega_0$.
As shown in Table~\ref{table:atomicpar}, our computed $K$ values for the relevant transitions in nickel HCIs are about $2$, which is higher than most of the current optical clocks.
For example~\cite{Flambaum2009}, Al$^{+}$ has $K= 0.008$. Out of $\sim 10$ species currently used in the optical clock community, only the heavy  Yb$^+$ and Hg$^{+}$ ions have $|K|>2$
~\cite{Flambaum2009}. 
Therefore, we expect that, even with their initial predicted accuracy of $10^{-19}$, the quantum clocks based on the relatively light Ni HCIs will have greater potential for exploring new physics than most of the current atomic clocks.
Recently, an improved constraint of  $\dot{\alpha}/\alpha=1.0(1.1)\times 10^{-18}/$year was reported based on the comparison of the $^2S_{1/2} (F=0) - {^2D}_{3/2} (F=2)$ ($E$2, $K=1.00$) and the $^2S_{1/2} (F=0) - {^2F}_{7/2} (F=3)$ ($E$3, $K=-5.95$) transition of $^{171}$Yb$^+$ clock~\cite{Lange2021}. 
The constraint on the temporal variation of $\alpha$ is expected to be further improved  by comparing two clocks based on the $E$2 transition of Ni$^{12+}$ and the $E$3 transition of Yb$^{+}$, because of its larger $K$ value and smaller projected systematic and statistical uncertainties of the $E$2 transition in Ni$^{12+}$ than those of the $E$2 transition in Yb$^+$.

\citet{Nandy2013} determined the sensitivity coefficient to the $\alpha$-variation for the ${^2\!P_{1/2}} - {^2\!P_{3/2}}$ transition in Ni$^{11+}$ ion. 
In their work, the transition rate and the lifetime of the 
${^2\!P_{1/2}}$ state were calculated using the relativistic coupled-cluster (RCC) method.
Yu and Sahoo~\cite{Yu2016,Yu2018} calculated some atomic parameters for the ${^2\! P_{3/2}} - {^2\!P_{1/2}}$ transition in Ni$^{15+}$ and the ${^3\!P_0} - {^3\!P_2}$ transition in Ni$^{12+}$ with the RCC and MCDHF methods.
Their results are also listed in Table~\ref{table:atomicpar} for comparison.
For lines $a$, $d$, and $e$, our calculated values agree well with other theoretical results~\cite{Nandy2013,Yu2016,Yu2018}, except for a factor of 3 difference for the sensitivity coefficient $q$ of line $d$. There is also a factor of 6 difference in the value of the  $Q$-factor  of line $e$, for which we traced back to the trivial factor of $2\pi$ missing in the linewidth definition in Ref.~\cite{Yu2018}.

Previous theoretical work on nickel HCIs focuses on atomic properties relevant to  the  emission from the solar, astrophysical, and laboratory plasmas.
In Tables~\ref{table:ComparisonTransitionrate} and~\ref{table:ComparisonLifetime}, we present a comparison with the literature values for the spontaneous decay rates and lifetimes. Overall, our MCDHF values agree well with the results from other theoretical methods, such as the RCC method and the multi-reference M{\o}ller-Plesset perturbation theory.
Moreover, the lifetimes of the $^2\!P_{1/2}$ state in Ni$^{11+}$ ion, the $^3\!P_{1}$ state in Ni$^{12+}$ ion, and the $^2\!P_{3/2}$ state in Ni$^{15+}$ ion were measured at the heavy-ion storage ring~\cite{Trabert2004-1, Trabert2004-2, Trabert2009}. We found a good agreement between theory and experiment.

\begin{table*}
   \caption{\label{table:ComparisonTransitionrate}
   Spontaneous emission rates in Ni HCIs, in s$^{-1}$. 
   }
    \begin{tabular}{ccc}
 \hline
\hline
  Line & This work & Other theory       \\
  \hline
  $a$    & 238(2)     & 235~\cite{Bilal2017}, 260~\cite{Zanna2016}, 236.31(3)~\cite{Nandy2013}, 237~\cite{Kaufman1986}, 213.1~\cite{HUANG1983}   \\
  $b$    & 157(1)   & 154~\cite{Yu2018},156.9~\cite{Fischer2010}, 157.4~\cite{Biemont1986,BHATIA1998}, 157~\cite{Kaufman1986,Mendoza1983}, 156~\cite{MALVILLE1965}            \\
  $c$    & 56.1(5)    & 56.08~\cite{Jonsson2016}, 57~\cite{Zanna2014}, 52.7~\cite{Landi2012}, 56.45~\cite{Fischer2010}, 56.42~\cite{Ishikawa2001},  56.5~\cite{Kaufman1986}, 54.66~\cite{HUANG1985}, 56~\cite{MALVILLE1965}        \\
  $d$    & 193(2)   & 192.2~\cite{EKMAN2018}, 190.99~\cite{Yu2016}   \\
  $e$    & 0.037(4)   & 0.034~\cite{Yu2018}, 0.03622~\cite{Fischer2010} 0.037~\cite{BHATIA1998},0.03702~\cite{Biemont1986}, 0.0355~\cite{Mendoza1983}, 0.048~\cite{MALVILLE1965}  \\
  $f$    & 0.030(1)   & 0.03~\cite{Jonsson2016},0.029~\cite{Landi2012}, 0.03044~\cite{Fischer2010}, 0.0157~\cite{Ishikawa2001}, 0.031~\cite{HUANG1985}  0.028~\cite{MALVILLE1965} \\
\hline
\hline
\end{tabular}
\end{table*}

\begin{table*}
   \caption{\label{table:ComparisonLifetime}
    Lifetimes (in ms) of upper clock states in Ni$^{11+}$, Ni$^{12+}$, Ni$^{14+}$, and Ni$^{15+}$. Numbers in square brackets stand for the powers of 10, {\it i.e.}, $x[y] \equiv x \times 10^y$.
    }

    \begin{tabular}{lcccc}
     \hline
        \hline
  Ion     & State                  & This work  & Other theory                                                          &Experiment       \\\hline
Ni$^{11+}$& ${^2\!P_{1/2}}$& 4.2(1)      & 4.25~\cite{Bilal2017}, 4.23(2)~\cite{Nandy2013}                  &4.166(60)~\cite{Trabert2004-1}        \\
Ni$^{12+}$& $^3\!P_{1}$            & 6.3(1)      & 6.5~\cite{Yu2018}, 6.55~\cite{Nazir2017}, 6.59~\cite{Nazir2017}   &7.3(2)$^*$, 6.50(15)$^{**}$~\cite{Trabert2004-2}  \\
Ni$^{12+}$& $^3\!P_{0}$            & 21(3)[3]    & 22.1[3]~\cite{Nazir2017}, 19.5[3]~\cite{Nazir2017}, 19[3]~\cite{Yu2018}    &                 \\
Ni$^{14+}$& $^3\!P_{1}$            & 17.8(2)     & 17.8~\cite{Jonsson2016}, 17.7~\cite{Ishikawa2001}                    &                 \\
Ni$^{14+}$& $^3\!P_{2}$            & 44(1)       & 44.6~\cite{Jonsson2016}, 45.1~\cite{Ishikawa2001}                    &                 \\
Ni$^{15+}$& ${^2\!P_{3/2}}$ & 5.2(1)     & 5.2~\cite{EKMAN2018}, 5~\cite{Yu2016}, 5.184~\cite{Santana2009}   &5.90(1)$^*$, 5.27(7)$^{**}$~\cite{Trabert2009}  \\
        \hline
        \hline
\multicolumn{5}{l} {* Single exponential evaluation} \\
\multicolumn{5}{l} {** Multi-exponential evaluation} \\
        \end{tabular}
\end{table*}

\subsection{Computational uncertainties}

The computational uncertainties in our work include the neglected correlation contributions, such as the triple- and quadruple-electron excitations involving the $1s$ orbital. The upper limit on these effects was estimated from the double excitations of the core orbitals in the single-reference configuration RCI calculations. The ``truncation” uncertainties due to the finite number of virtual orbitals were evaluated based on the convergence trends in the above-mentioned three steps. For the wavelengths, all the uncertainties were summed together in quadrature. For the $M$1 transition rates, in addition to these uncertainties, we also included the frequency-dependent Breit interaction contribution as another source of error.  For the $E$2 transition rates, the difference in results between the Babushkin and Coulomb gauges~\cite{Grant1974} were treated as an additional contribution to the combined theoretical uncertainty.

\section{CONCLUSIONS}

To reiterate, the quantum clockwork we explored here provides an intriguing possibility for achieving  high accuracy on multiple transitions in HCIs of the same element.
Our strategy offers an important flexibility in the pursuit of multiple candidate clock transitions.
Particularly, the $E$2 transition in $^{61}$Ni$^{12+}$ has projected fractional uncertainty  $10^{-19}$.
We demonstrated the key experimental capabilities of using our SW-EBIT facility to generate and extract
Ni$^{11+}$, Ni$^{12+}$, Ni$^{14+}$, and Ni$^{15+}$ ions.
We measured the wavelengths of four $M$1 and one $E$2 clock transitions in these ions with the uncertainties of about 2~pm. The measured wavelengths establish an important reference for precision laser spectroscopy in future clock transition measurements.
We also calculated spectroscopic properties of the relevant $M$1 and $E$2 clock transitions. 
The calculated wavelengths are consistent with our experimental results and with previous determinations.
The calculated properties indicate that these ions are suitable for precision quantum metrology and for exploring new physics beyond the standard model of particle physics.


\begin{acknowledgments}
The authors thank Xin Tong, Jos\'e R. Crespo L\'opez-Urrutia, and Yan-Mei Yu for helps and for fruitful discussions.
This work was supported by the Strategic Priority Research Program of the Chinese Academy of Sciences (Grant No. XDB21030300),
the National Natural Science Foundation of China (Grant Nos. 11934014, 11622434, 11974382, 11604369, 11974080, 11704398, and 11874090),
the National Key Research and Development Program of China under Grant No. 2017YFA0304402,
the CAS Youth Innovation Promotion Association (Grant Nos. Y201963 and 2018364),
the Hubei Province Science Fund for Distinguished Young Scholars (Grant No. 2017CFA040),
and the K. C. Wong Education Foundation (Grant No. GJTD-2019-15). ZCY was supported by NSERC of Canada.
Work of A.D. was supported in part by the U.S. National Science Foundation.
\end{acknowledgments}

\end{document}